\documentclass[10pt,journal,compsoc]{IEEEtran}

\ifCLASSOPTIONcompsoc
  \usepackage[nocompress]{cite}
\else
  \usepackage{cite}
\fi

\ifCLASSINFOpdf
\else
\fi

\usepackage{xcolor}
\usepackage{graphicx}
\usepackage{multirow}
\usepackage{url}
\usepackage{rotating}
\usepackage{textcomp}
\usepackage{amsmath}

\begin{document}

\title{SEC-NoSQL: Towards Implementing High Performance Security-as-a-Service for NoSQL Databases}

\author{G.~Dumindu~Samaraweera\IEEEmembership{}        
        and~J.~Morris~Chang~\IEEEmembership{}
\IEEEcompsocitemizethanks{\IEEEcompsocthanksitem G.D. Samaraweera is with the Department of Electrical Engineering, University of South Florida, 4202 E. Fowler Avenue, Tampa, FL 33620.\protect\\
E-mail: samaraweera@mail.usf.edu
\IEEEcompsocthanksitem J.M. Chang is with the Department of Electrical Engineering, University of South Florida, 4202 E. Fowler Avenue, Tampa, FL 33620.
\protect\\ E-mail: chang5@usf.edu
}
\thanks{ }}

%
%

\markboth{ }%
{Samaraweera \MakeLowercase{\textit{et al.}}: Towards Implementing Security-as-a-Service for NoSQL Databases}

\IEEEtitleabstractindextext{%
\begin{abstract}
During the last few years, the explosion of Big Data has prompted cloud infrastructures to provide cloud-based database services as cost effective, efficient and scalable solutions to store and process large volume of data. Hence, NoSQL databases became more and more popular because of their inherent features of better performance and high scalability compared to other relational databases. However, with this deployment architecture where the information is stored in a public cloud, protection against the sensitive data is still being a major concern. Since the data owner does not have the full control over his sensitive data in a cloud-based database solution, many organizations are reluctant to move forward with Database-as-a-Service (DBaaS) solutions. Some of the recent work addressed this issue by introducing additional layers to provide encryption mechanisms to encrypt data, however, these approaches are more application specific and they need to be properly evaluated to ensure whether they can achieve high performance with the scalability when it comes to large volume of data in a cloud-based production environment. This paper proposes a practical system design and implementation to provide Security-as-a-Service for NoSQL databases (SEC-NoSQL) while supporting the execution of query over encrypted data with guaranteed level of system performance. Several different models of implementations are proposed, and their performance is evaluated using YCSB benchmark considering large number of clients processing simultaneously. Experimental results show that our design fits well on encrypted data while maintaining the high performance and scalability. Moreover, to deploy our solution as a cloud-based service, a practical guide establishing Service Level Agreement (SLA) is also included.
\end{abstract}

\begin{IEEEkeywords}
Big Data, Database Systems, NoSQL, Security, Performance, Security as a Service.
\end{IEEEkeywords}}

\maketitle

\IEEEdisplaynontitleabstractindextext

%
\IEEEpeerreviewmaketitle

\IEEEraisesectionheading{\section{Introduction}\label{sec:introduction}}

%
%
%
%

\IEEEPARstart{T}{he} concept of Big Data and cloud computing paradigm has empowered numerous applications in our day-to-day life and as a result, during the last few years it generated very large volume of data. This volume expansion of data with a fast pace imposes several challenges when storing and managing data. At the same time, it brings intensive needs of data access on database systems, making them to change the architecture from centralized mechanism to distributed nature while managing the stress of very large amount of data. Due to the nature of relational models and their structured data schemes, traditional relational database systems were unable to provide the demanding requirements of Big Data analytics. This brings the existence of NoSQL data models in which they bundled with many added advantages compared to relational databases including the support for unstructured data, high concurrency, low latency, high flexibility, high scalability and availability, and reduced operational and management cost which eventually help many organizations to provide better quality of service. There are different types of NoSQL data models that are actively being discussed and these can be categorized in to four basic types. 1) Key-Value Stores having a big Hash Table of keys and values (e.g. Riak, Amazon DynamoDB) 2) Document-Oriented Stores that can handle documents made up of tagged elements (e.g. MongoDB, CouchDB) 3) Column-Oriented Stores where each storage block contains data from only one column (e.g. Cassandra, HBase) 4) Graph Stores which can be a network database that uses edges and nodes to represent and store data (e.g. Neo4J, OrientDB).

With the evolving improvements in the cloud computing arena, distributed storage architecture has given the existence for Database-as-a-Service (DBaaS) solutions where data owners and data users can access the mass scale remote data storage. Many cloud vendors have given attractive service offerings through this model so that organizations can deploy their backends in cloud with low cost and guaranteed high availability. However, one of the major concerns in this model is, after storing data in the cloud infrastructure, data owner does not have any control over his data. Many cloud users concern about the privacy and security of their data on which cloud operators have the access. Researches have been carried out to address this issue in different dimensions but practicality of those solutions with respect to the demanding performance requirements of the databases (especially the NoSQL databases) in cloud paradigm while ensuring security, is still unresolved.

Despite the benefits of using NoSQL databases on cloud platform, ensuring Confidentiality, Integrity and Availability (CIA) of data along with the privacy of the data owner is one of the major concerns. As per a report \cite{CloudSecurityAlliance2013} published by Cloud Security Alliance (CSA), most of the NoSQL data stores still suffer from data protection strategies as they are more concerned on performance and scalability. On the other hand, it is much challenging for a database system to ensure complete security on a cloud-based architecture, as different clients may have different set of applications and each application can have completely different set of security requirements. In fact, they leave the security part for the client-side, to implement based on application specific requirements. 

The research on security-aware database systems based on different approaches have discussed for many years. However, with the increasing popularity of NoSQL databases, security-aware implementation of databases became more popular and a necessity. The one of the approaches to ensure security and privacy for the sensitive data on a database in a cloud environment, is to implement an efficient encryption mechanism either at the application level or on top of a middle-ware application. However, just encrypting the data arbitrarily in a database, prevents further querying and updating. This has motivated some research on providing restricted set of tasks like equality checks and range queries on encrypted data in a database system. CryptDB \cite{popa2011cryptdb} is one of such systems developed for a relational database, utilizes a middle-ware application to rewrite the original query issued by the client, making them executed on the encrypted data at the database level. This model has given the diversity to implement security on cloud-based database systems. Nevertheless, application of cryptographic techniques is always aligned with an additional cost which affects the performance. Thus, it is a challenging task for a NoSQL database itself to ensure security without compromising the performance. In contrast to the relational databases, research on security-aware NoSQL databases is poorly addressed in the past. The efforts carried out in the past have not evaluated on increasing/different heavy workloads in par with the realistic cloud environment(s). On the other hand, some of them suggested considerable amount of changes either on client-side application or in the database itself. For an example \cite{yuan2016building} requires architectural changes at database while \cite{macedo2017practical} requires to employ specially designed 'Crypto Worker' both at the client-side and server-side. 

This paper addresses these issues and proposes a new practical design and implements a Security-as-a-Service (SECaaS) solution for NoSQL databases, which can provide the functionality of query over encrypted NoSQL database. Our main intention is to provide a high performance solution that can withstand the additional cost incurred when implementing cryptographic primitives on NoSQL database systems. The solution provides an extensive platform for the database users to configure their application specific security requirements without really touching the client application or database engine while optimizing the system performance, which can be simply deployed on a cloud infrastructure. With that, our design offers all three components of CIA triad. In a nutshell, our main contributions are as follows.

\begin{itemize}
	\item Design and implement a Security-as-a-Service (SECaaS) solution for NoSQL databases (called SEC-NoSQL) which has the fully functional capability to query over encrypted data with guaranteed performance even in increasing workloads. Particularly, our system can be adopted in the existing environments without doing any architectural changes to the client-side code or database engine and it can be deployed as Secure-DBaaS solution on hybrid cloud environment.
	
	\item An extensive performance evaluation of the different system models based on Cassandra NoSQL database and Yahoo! Cloud Serving Benchmark (YCSB).
	
	\item Provide a granular specification which enables implementing Service Level Agreements (SLA) (for cloud service providers) based on the experimental results by considering the deployment strategies of SEC-NoSQL.
\end{itemize}

The rest of the paper is structured as follows. Section 2 discusses the relevant related previous work of security-aware database implementations. The section 3 discusses about the threat model of the proposed design. In section 4 different cryptographic schemes and their implementations related to SEC-NoSQL are introduced. Section 5 describes the architecture of the system. Section 6 gives a detailed discussion on system implementation and experimental evaluation. Finally, paper concludes section 7 giving further observations and future work.

\section{Related Work}
Practical issues when implementing security-aware databases in production environment has been in discussion for more than a decade now. Most of these discussions were centered around relational database systems on a security stand point. However, with the increasing demand for a high scalable and unstructured database systems, the point where NoSQL databases has given the existence, security was not the major concern. Thus, NoSQL database systems were initially shipped with poor set of security implementations or lack of security. In 2011, Okman et al. pointed out main security features and problems \cite{okman2011security} in two of the most popular NoSQL databases, Cassandra and MongoDB which are still in the list of top ten popular NoSQL database systems \cite{DB-Engines2018}. In a different prospect Soria-Comas et al. \cite{soria2016big} discussed the challenges to the privacy principles and models and evaluated how well main privacy models meet the requirements of Big Data. All these previous works have concluded the importance of having security-aware NoSQL database systems.

In a context where security-aware relational databases, CryptDB \cite{popa2011cryptdb} introduced a well-structured design for a secure database system. The main idea was, client rewrites the original SQL query in a trusted vicinity and sends it to the database so that the database system can execute them over encrypted data located in an untrusted environment. Their design was bundled with layered architecture of encryption schemes, which enables execution of equality checks, order comparisons, aggregates, and joins. This idea has given the momentum to implement different security-aware database systems. In a different CryptDB based framework, Monomi \cite{tu2013processing} was able to perform analytical queries over encrypted data by splitting the query execution between server and client. In an another approach, L-EncDB \cite{li2015encdb}, implemented a light weight encryption mechanism designed for databases which enables SQL based queries over encrypted data while preserving the same length of the data structure.

Several different approaches have also been proposed to implement security-aware NoSQL data stores. BigSecret \cite{pattuk2013bigsecret} is a framework that enables secure outsourcing and processing of encrypted data over key-value stores where indexes are encoded in a way that allow comparisons and range queries. In addition, L-EncDB also has their variant for the support on NoSQL databases. In another variant of CryptDB, Arx \cite{poddar2016arx}, is a database system that encrypts the data with stronger security guarantees. Instead of embedding the computation into special encryption schemes as in CryptDB, Arx embeds the computation into data structures, which it builds on top of traditional encryption schemes. In an another quietly different approach, Yuan et al. \cite{yuan2016building} proposed an encrypted, distributed, and searchable key-value store along with a secure data partition algorithm that distributes encrypted data evenly across a cluster of nodes. In SecureNoSQL, Ahmadian et al. \cite{ahmadian2017securenosql} looked in to the aspects of ensuring both confidentiality and integrity of data on a document-store NoSQL data model. However, their design consists of an additional 'Security Plan' attached to each document submitted by the data owner; thus, original client application needs to be modified. Macedo et al. \cite{macedo2017practical} also presented a generic NoSQL framework and set of libraries supporting data processing and cryptographic techniques that can be used with existing NoSQL engines. However, most of the previous studies have been designed only focusing on specific set of protection schemes or techniques, but much attention has not given to practically implement the solution by ensuring security without sacrificing the functionality and performance on NoSQL databases in a production environment with high volume of data. Particularly, our work, SEC-NoSQL can provide query over encrypted data in a distributed fashion ensuring both confidentiality and integrity by maintaining a guaranteed level of performance. In addition, it can be simply adopted to existing environments without doing any architectural changes; thus, client-side applications and database engines are not really modified.

In addition to the embedded security offerings, couple of different (general) security-as-a-service models have also been proposed over the past to protect the cloud infrastructure from security threats. Varadharajan et al. \cite{varadharajan2014security} proposed a security service model for cloud providers to protect their own infrastructure, enabling the flexibly for tenants to determine how much control they wish to have over tenant's virtual machines. In \cite{hawedi2018security} Hawedi et al. proposed another security architecture that offers elastic security for cloud tenants that can be deployed on hybrid clouds. In a different context, to mitigate the vulnerabilities at the application layer, Chen et al. \cite{chen2018security} proposed a security architecture that offers security service throughout software development cycle. However, all of these solutions are trying to address the external threats (either using infrastructure layer VMs, network layer traffic or application layer). Yet, none of these solutions are addressing the internal threats (discussed in Section \ref{sec:Threat Model}) originated at the database level. For the best of our knowledge this is the first approach to implement security-as-a-service for cloud-based database services.

Apart from the security concerns, significant number of studies were carried out to evaluate the performance measures of NoSQL databases. Klein et al. \cite{klein2015performance} has studied the performance of three NoSQL databases namely MongoDB, Cassandra and Riak. Their experiments were done in a production scale environment with set of electronic health-care records by considering the Consistency, Availability, and Partition tolerance (CAP) theorem \cite{brewer2000towards} trade-offs and evaluated using the popular Yahoo! Cloud Serving Benchmark (YCSB) developed by Cooper et al. \cite{cooper2010benchmarking}. Based on their research, Cassandra provided the best overall performance, at the same time it also delivered the highest average latencies. In a different context, Waage et al. \cite{waage2014benchmarking} evaluated the performance overhead of encrypting data on Cassandra and HBase, again using the YCSB. And their results show that turning on the encryption reduces the overall throughput approximately 40\% on average for single read/write operations and it was only 10\% of reduction when it comes to range scans.

\section{Security-as-a-Service Threat Model} \label{sec:Threat Model}
The threat model of our system is focuses on honest-but-curious passive attacks which is consistent with related works on encrypted database systems \cite{popa2011cryptdb}, \cite{ahmadian2017securenosql}. Thus, the attacker is assumed to be passive, acts in an honest manner, trying to learn some confidential information but does not change any queries issued by the proxy or query results. Based on the information available to the cloud server, we consider two adversarial threat models with different attack capabilities as below.

\begin{figure}[ht]
	\centering
	\includegraphics[width=0.9\columnwidth]{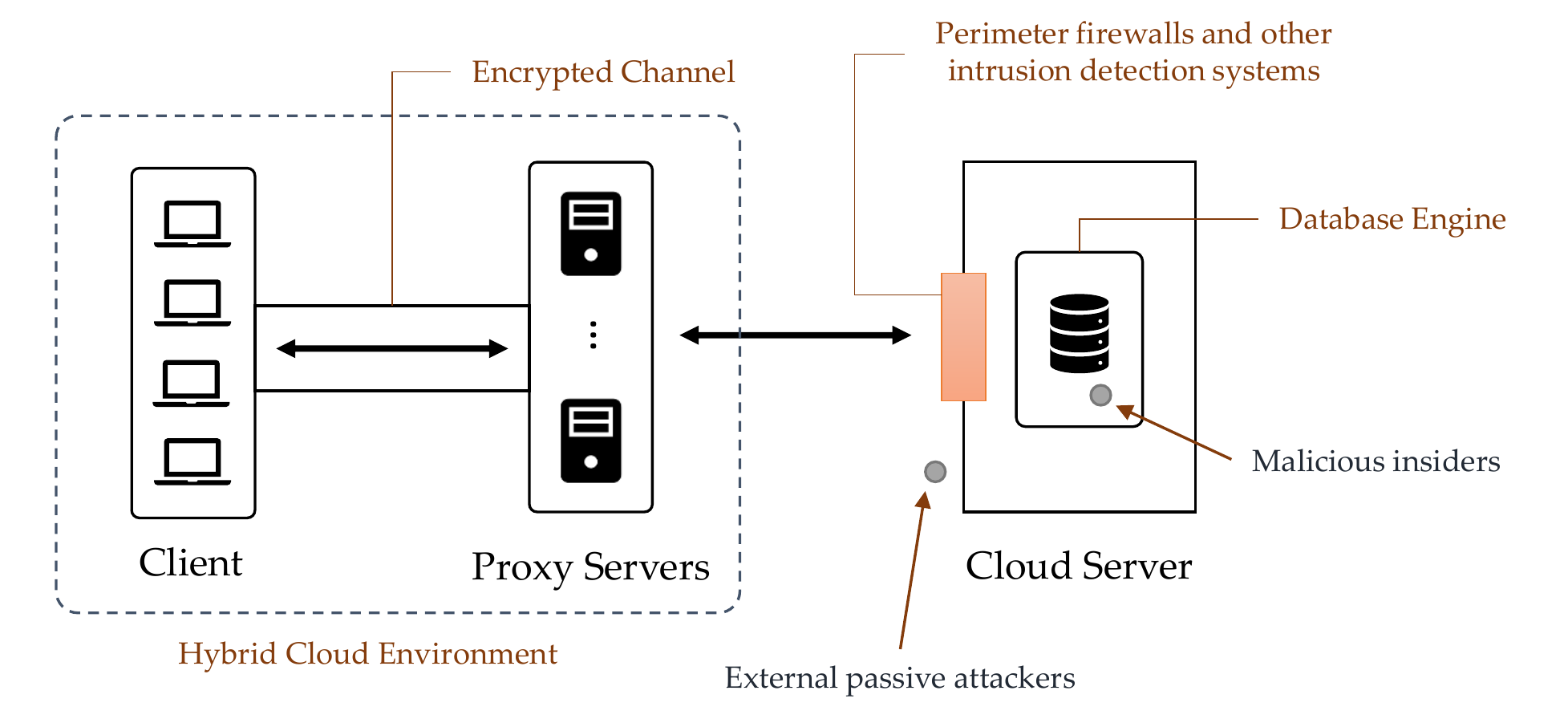}
	\caption{Overview of the Threat Model.}
	\label{fig:threat_model}
\end{figure}

\begin{itemize}
	\item \textbf{Malicious Insider Attacks:} \\
	With the increased use of cloud computing infrastructures to deploy database services, the cloud system administrators and database administrators typically have access to privilege domains. Hence, this threat is increasingly important to address; and in our design, we focus on both confidentiality and integrity protection mechanisms to overcome this threat. The application of encryption mechanisms before data get stored in the cloud database server ensures the confidentiality of data from such threats. Moreover, integrating a verification mechanism to the proxy ensures the data integrity. 
	\\
	\item \textbf{External Passive Attacks:} \\
	While malicious insider attacks are originated within the cloud database, there can be a possible attacker who compromises the database server externally and observes its operations including queries issued to the database and how they access data. Observations of query evaluation can damage the encrypted data content specially on deterministic and order-preserving encryption mechanisms \cite{grubbs2016breaking}. Hence it is important to implement a strong security primitive like Oblivious Random Access Machine (ORAM) in order to mitigate such access pattern attacks. However, in this work we consider this external attack is limited to the server-side (cloud) where cloud service providers have vested interest on protecting their resources by using encrypted communication channels, rigid firewalls and other intrusion detection systems.
\end{itemize}

The Fig. \ref{fig:threat_model} shows the overview of the threat model along with the possible abstraction of the attack.

\section{Ensuring Security while Processing Queries over Encrypted Data}
To facilitate query over encrypted records on a database, it is required to implement cryptographic techniques in a way such that it allows to perform basic read/write operations over encrypted data. Different encryption algorithms provide different privacy guarantees. A more secure encryption algorithm can ensure the indistinguishability against a powerful adversary while some schemes allow different operations to be performed over encrypted data. In this section we explore the different security schemes and the applicability towards implementing security-aware NoSQL, based on the schemes that are widely used in the existing literature. Moreover, our research evaluations are specifically focused on Apache Cassandra \cite{Foundation2008a}, a top leading wide-column store NoSQL database. Thus, these cryptographic primitives are explained according to the Cassandra Query Language (CQL) implementation specific point of view. However, these implementations can be extended the same way as shown under the sub section \ref{subsec:Cryptographic Implementations} into other different NoSQL database systems.

\subsection{Standard/Random (RND) Encryption}
This is a probabilistic encryption scheme which provides the maximum security where two equal values are mapped to different ciphertexts every time, with high probability. This scheme provides robust level of security which is computationally hard for an adversary to derive significant information from a ciphertext. Due to this robustness, it is also hard to perform meaningful and realistic computations on it and query over encrypted data. In our design we implemented RND using Advanced Encryption Standard (AES) \cite{miller2009advanced} with a random Initialization Vector (IV).

\subsection{Deterministic (DET) Encryption}
In this scheme when the encryption is performed, same plaintext deterministically produces the same ciphertext every time, thereby allowing to perform equality checks on the encrypted data. Intuitively, for the encryption algorithm DET and two messages $msg1$ and $msg2$ where $msg1 = msg2$ then, $DET(key,msg1) = DET(key, msg2)$ for all $msg1$ and $msg2$. Therefore, this scheme allows the encrypted database to perform $SELECT$ queries with equality checks, equality JOINs, GROUPBY, COUNT and DISTINCT clauses. In cryptographic terms, DET should be a pseudo-random permutation and in our implementation, AES with fixed IV is used. However, comparatively DET is less secure than RND since it could reveal the information of plaintext-ciphertext pair.

\subsection{Order-preserving Encryption (OPE)}
OPE is an encryption scheme which satisfies inequality where two messages, $msg1$ and $msg2$ with $msg1 > msg2$ then $OPE(key,msg1) > OPE(key, msg2)$ for any secret key, $key$. This allows to determine order relations on the encrypted data without revealing the original data itself. However, since it could reveal the order in the original content, OPE is considered as a weaker encryption scheme compared to DET. Moreover, it is worth to note that inference attacks with higher success rate can be performed on OPE encrypted columns and an attacker can extract the sensitive information without much effort as pointed in \cite{naveed2015inference}. Therefore, it is important to carefully analyze the tradeoffs between security and optimized query operations on encrypted data, before implementing OPE encryption schemes in sensitive data records.

\subsection{Homomorphic Encryption (HOM)}
This is again a probabilistic encryption scheme having the capabilities to perform computations over encrypted data. Fully homomorphic encryption enables the computation on encrypted data with operations such as addition and multiplication. To perform an operation like addition in SQL data field (for e.g. $SET amount = amount + 1)$, homomorphic encryption can sustain where multiplying the encryptions of two values results in encryptions of sum of the two values, $HOM(key, x) \cdot HOM(key, y) = HOM(key, x+y)$.

\subsection{Cryptographic Implementations in SEC-NoSQL}  \label{subsec:Cryptographic Implementations}
For the performance evaluations, our implementation is focused on Cassandra NoSQL database; however, the same architecture can be extended to other NoSQL models such as MongoDB. Even though, most of the query representations of CQL (query language of Cassandra), is similar to general SQL implementations in any other relational database system, there are few slight changes. Because of the non-relational model, most of the NoSQL databases including Cassandra, MongoDB do not support JOIN queries. Further, in order to perform SELECT queries on columns, Cassandra requires that column to be indexed (if it is not a member of the Partition Key) \cite{DataStax2018}. Moreover, Partition Key (first element of the PRIMARY KEY) has a special use in Cassandra beyond showing the uniqueness of the record in the database. Thus, it is also important that each query in Cassandra requires to specify the partition key. Table \ref{tab:comparison cql} shows a general comparison between how different operations can be done using general SQL queries and CQL queries.

\begin{table}[hbtp]
	\centering
	\caption{General Query Comparison Between SQL and CQL.}
	\includegraphics[width=0.99\linewidth]{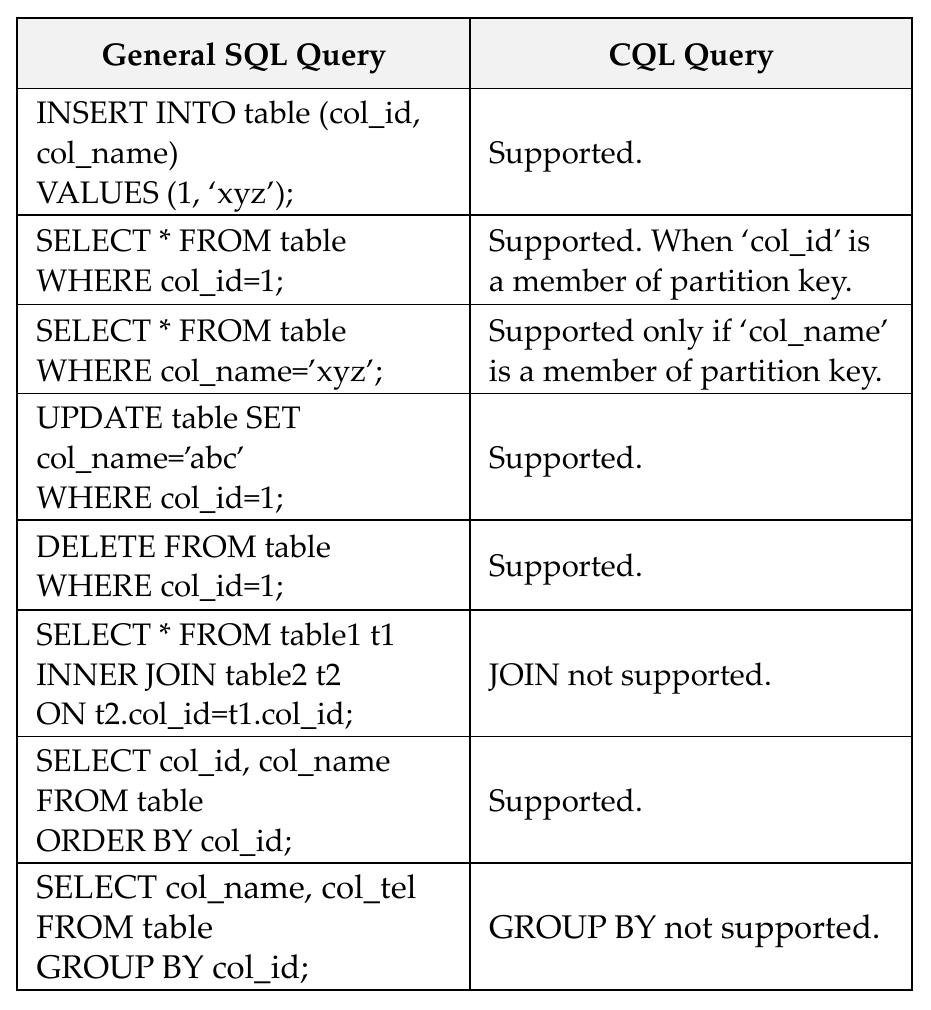}	
	\label{tab:comparison cql}
\end{table}

In a database system, CRUD operations are the four basic functions of persistent storage. CRUD stands for Create, Read, Update and Delete. Based on our study, to implement a security-aware encrypted Cassandra database, we only require RND, DET and/or OPE encryption schemes. Thus, RND and DET encryption schemes have been implemented in our solution so that all CRUD operations can be performed to evaluate the performance of the system. We leave order relation queries as a potential future work. To that end, an example showing the query translation implemented in SEC-NoSQL is shown in Table \ref{tab:support for CRUD}. In terms of extending our work for other NoSQL models, it also compares the query representations with MongoDB query language. In the given example 'col\_id' is the key and 'col\_name' is the value. Typically, in SEC-NoSQL, column names and table names are anonymized to ensure privacy. As it is shown in the Table \ref{tab:support for CRUD}, when a record is written to a table (Insert/Update), Primary Key field ('col\_id') is encrypted with DET encryption scheme, allowing us to uniquely identify at the next time for a database read/update whereas other fields are encrypted using RND. When retrieving data from the database, first, 'col\_id' value given by the user's original query will be encrypted using DET. Thereafter, that encrypted value is used to search for a corresponding record against the encrypted database. Once it is found, results will be decrypted accordingly.

\begin{table}[hbtp]
	\centering
	\caption{SEC-NoSQL Support for CRUD Operations.}
	\includegraphics[width=0.99\linewidth]{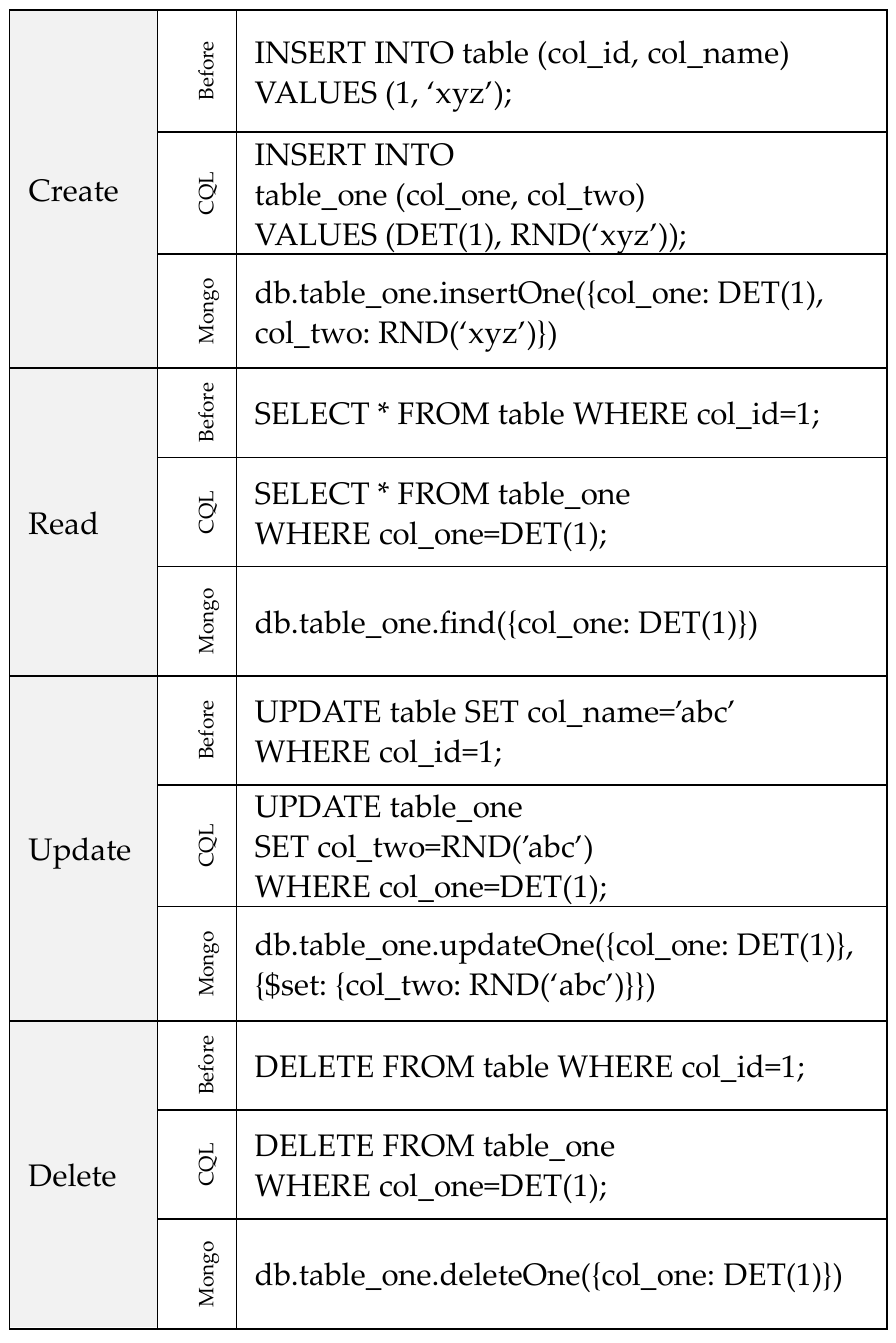}	
	\label{tab:support for CRUD}
\end{table}

To perform the query translation, encryption, decryption along with the integrity check, a middle-ware application (proxy) is introduced. The main advantage of having a proxy application at the middle (deployed on a hybrid cloud) is that we can adopt the SEC-NoSQL in to the existing production environments without having any architectural changes to the client-side or database engine. In the next section, we introduce detailed system design and architecture of the SEC-NoSQL.

\section{Design and Architecture of SEC-NoSQL}
The main objective of the proposed architecture is to implement a practical, security-aware NoSQL database solution that operates as a cloud-based service, with guaranteed level of performance and scalability. We have designed our solution in a way such that it can be fitted in to any type of NoSQL data store even though we have implemented and evaluated on Cassandra database cluster. Moreover, this design can be simply deployed as SECaaS solution for NoSQL databases which is discussed in \ref{sec:Implementation and Evaluation}. In addition, existing users in a production environment who requires backward compatibility, can continue to directly communicate with Cassandra without having the Proxy server(s). 

\subsection{Importance of having Hybrid Cloud Architecture}
\begin{figure}[ht]
	\centering
	\includegraphics[width=0.99\columnwidth]{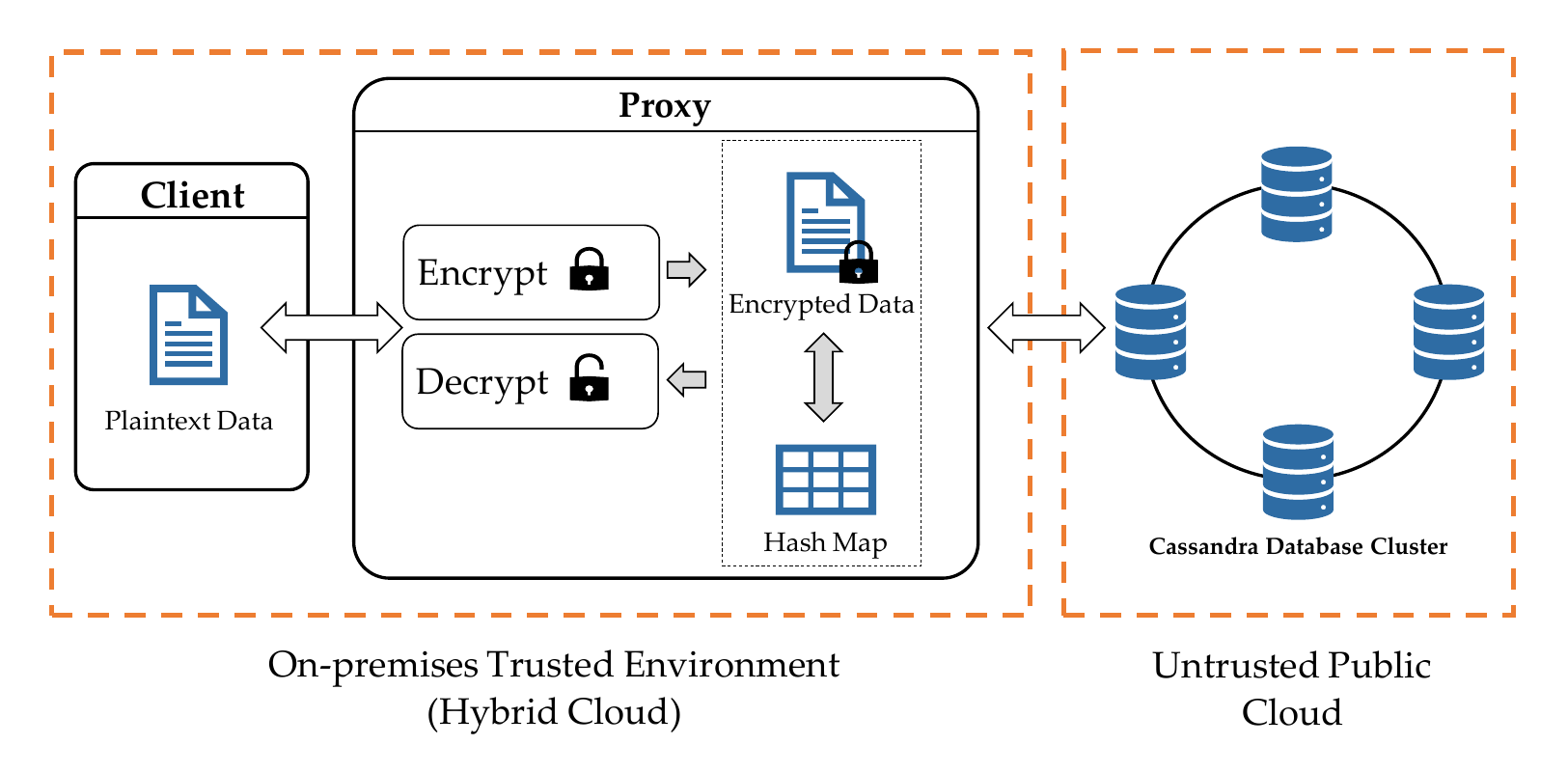}
	\caption{{Basic Architecture of SEC-NoSQL.}}
	\label{fig:basic_architecture}
\end{figure}

Cloud based architecture has been part of the data management systems for several years now. However, many organizations are realizing that cloud service providers alone cannot deliver all the necessary tools to efficiently manage heterogeneous environments, especially in terms of data security. To meet this challenge, there is a growing interest in adopting hybrid data model to data infrastructure where some data may be on-premise while rest of the data on the cloud. In general, a hybrid cloud refers to a cloud environment made up of a mixture of on-premises private cloud resources combined with third-party public cloud resources that use some kind of orchestration between them. The main advantage of employing such hybrid cloud model is that it allows workloads and data to move between private and public clouds giving greater flexibility. In our security architecture, we introduce this hybrid cloud model to deploy the proxy server and a high level view of this architecture is depicted in Fig. \ref{fig:basic_architecture}.

\subsection{SEC-NoSQL Design}
In addition to the details discussed in the previous sections, in our design, clients only communicate with the Proxy server as they would do in a scenario where storing or retrieving data directly from the database. At first, data owner issues a request to the Proxy for database schema creation. Proxy transform the schema by anonymizing the column names and execute it at the database. When a request appears at the Proxy for data write, after the encryption process, Proxy generates a hash value (HMAC) of that record using HMAC-SHA256 algorithm. This value then will be stored at a Hash Table maintained by Proxy. At an event of read request, Proxy retrieve the encrypted record from the database, generate the hash value and compare the corresponding entry in the Hash Table to ensure the data integrity. If the integrity check is passed, record then will be decrypted, and corresponding results will be sent back to the client.

A production ready cloud-based NoSQL database system is typically scalable across horizontal, termed as horizontal-scaling. With horizontal-scaling, it is often easier to scale the database dynamically by adding more machines to the existing cluster without specific downtime in the production environment. Particularly, Cassandra is a horizontally scalable database that can handle massive amount of data distributed across several commodity servers. In a high-level overview, Cassandra operates by dividing all data evenly around a cluster of nodes providing parallel data operation. Unlike other NoSQL implementations, Cassandra does not have master/slave architecture therefore all peers are capable of receiving requests \cite{DataStax2018a}. Based on these properties of Cassandra, 4 different secure implementation models have been proposed to evaluate the performance metrics of our design.

\subsection{No Encryption Model (NoEnc)}
\begin{figure}[ht]
	\centering
	\includegraphics[width=0.7\columnwidth]{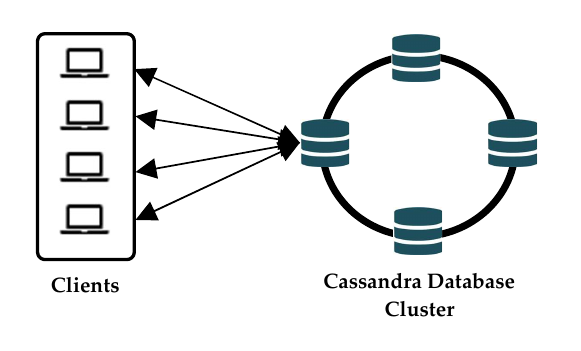}
	\caption{{NoENC Model where all clients are directly connected to the Database cluster.}}
	\label{fig:no proxy}
\end{figure}

This is the base model that we are using to evaluate the performance overhead of the other models thus, none of the encryption schemes are employed. In this model client directly connects to one of the nodes in the Cassandra cluster (no proxy involved in this setup). Client machine is enabled with YCSB application and generate an increasing workload by changing the number of client threads at different evaluation stages. Fig. \ref{fig:no proxy} represents the NoEnc model.

\subsection{Encryption Model 1 (EncM1)}
\begin{figure}[ht]
	\centering
	\includegraphics[width=0.7\columnwidth]{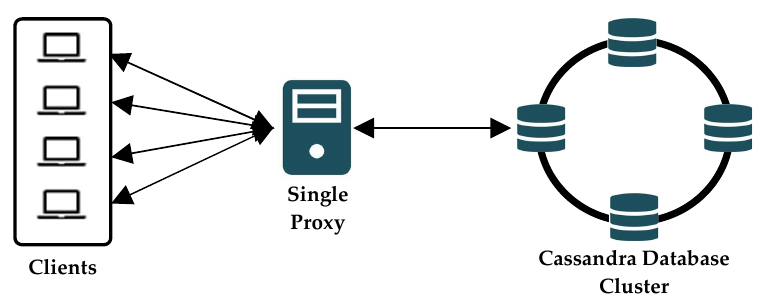}
	\caption{{EncM1 Model with single Proxy.}}
	\label{fig:encm1}
\end{figure}

Our first approach that is enabled with cryptographic schemes is named as EncM1. As shown in Fig. \ref{fig:encm1}, this model is equipped with a separate dedicated proxy server at the middle of the communication channel.

Due to this additional interface at the middle, an extra performance cost compared to NoEnc model is expected. In contrast to the several earlier approaches as literature suggested, we believe that having a dedicated proxy server (deployed over hybrid cloud) is more realistic in a production environment, than an add-on application at client-side, to withstand the performance overhead.

\subsection{Encryption Model 2 (EncM2)}
\begin{figure}[ht]
	\centering
	\includegraphics[width=0.7\columnwidth]{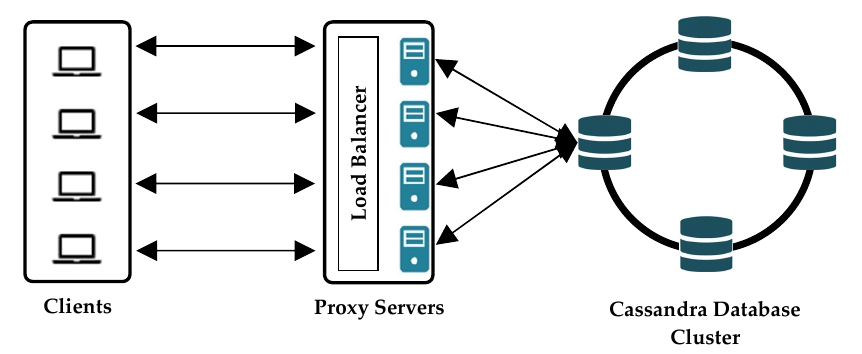}
	\caption{{EncM2 Model with several Proxy servers.}}
	\label{fig:encm2}
\end{figure}

With an increasing number of high volume client workloads, in practice, single proxy server might experience performance bottleneck due to network congestion and resource limitations. In this model we employ several proxy servers with an integrated load balancing mechanism. As it shows in Fig. \ref{fig:encm2}, each proxy server works parallelly to improve the overall throughput.

\subsection{Encryption Model 3 (EncM3)}
\begin{figure}[ht]
	\centering
	\includegraphics[width=0.7\columnwidth]{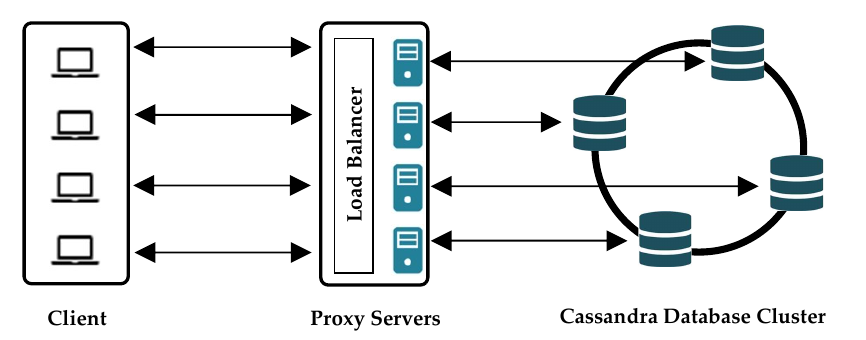}
	\caption{{EncM3 Model with several Proxy servers.}}
	\label{fig:encm3}
\end{figure}

Cassandra has a nice salient property that every node in a cluster can work as a coordinator to connect and accept database connections. Once connected, coordinator distributes the workload accordingly among all active nodes in the cluster. We employ this feature and designed the EncM3 model which is shown in Fig. \ref{fig:encm3}.

\section{Implementation \& Experimental Evaluation} \label{sec:Implementation and Evaluation}
\subsection{System Implementation}
SEC-NoSQL is implemented on Apache Cassandra, originally devised by Laskshman et al. \cite{Foundation2008a} at Facebook and later it was released as an open-source distributed database. Cassandra was inspired by Google BigTable and Amazon DynamoDB, and it is one of the most popular NoSQL databases in today because of its robust and salient features, as discussed in previous sections.

We have implemented our solution on a high-end environment (Intel Core i7-7820X/ 80GB) on top of virtual machines (VMs). Each component/module of the system is deployed on a separate VM with specific resource allocations managed by VMware Workstation 12. To make it a realistic environment, inter VM communication has been divided in to two virtual networks where Client machine and Proxy servers are residing in a one local network while Cassandra database servers are allocated to the other network. To manage the network level name resolutions, we employed a separate Domain Name Server (DNS) running on top of Ubuntu 16.04.

Moreover, we deployed our Cassandra 3.11 database cluster on top of Cloudera CDH 5 (Cloudera Distribution Hadoop) which is the most competed, tested and widely deployed open source distribution of Apache Hadoop, built specifically to meet enterprise demands \cite{Clodera2018}. To measure the performance of our design in relation with the Big Data analytics that focuses on data warehousing and data lake use cases, we integrated CDH to our solution. With this, we believe that our solution can be expanded as a service in to modern web, mobile and IoT possible use cases that supports analytics on hot data. On the other hand, Cloudera Manger (a single interface tool bundled with CDH to monitor and manage Big Data infrastructure) simplify the process of managing nodes by providing cluster-wide real-time view of nodes and services which in turn helps us to do horizontal-scaling on Cassandra.

Proxy server in our design plays an important role. It is basically a server application (implemented on Java) that listens for incoming request from client, then makes necessary query translations based on cryptographic implementations and finally execute the query against encrypted database and return the query execution results back to the client. We have implemented the RND and DET schemes using AES-128 in cipher block chaining mode (CBC). However, it is crucial to handle this task 1) parallelly, so that multiple requests can be handled at the same time 2) efficiently, by maintaining minimal congestion at incoming TCP port, to support a production environment that engaged with large volume of client requests. A separate execution thread (from a thread pool using an executor service) is allocated for each incoming client connection to process the requests parallelly. We implemented our solution using Netty 4.1 (a NIO based client-server framework) \cite{NettyProjectCommunity2016} as the network application framework with incorporating Google Protocol Buffers 3.5 \cite{Google2017} for the protocol support. This combination helps us to design an asynchronous event-driven network application server and client with high performance, stability and flexibility without a compromise.

Moreover, we have customized the YCSB 0.14.0 client in a way to incorporate network communication framework into it, so that YCSB can communicate with Proxy server(s) for the performance measures. In our test environment, we only have one client machine which in turn generate different workloads at different time using multiple client threads in order to simulate the high volume of requests coming from multiple clients. In addition, our implementation of SEC-NoSQL is fully stacked on virtual machines installed with Ubuntu 16.04 LTS (64-bit) and additional Client and Proxy applications were developed using Java 8. Client VM and each Proxy Server VM are assigned with 4core CPU and 2GB of memory while each Cassandra Node VM is assigned with 4core CPU and 4GB of memory by considering the minimum hardware requirements documented by Cassandra \cite{Cassandra2016}.

\subsection{Background Work for Experimental Evaluation}
For the evaluation of our proposed models against the performance, we created a Cassandra cluster with the replication factor assigned to four. Begin with every evaluation, total of 40,000 data records were generated using YCSB and pre-loaded in to the database. Furthermore, at client-side in YCSB, we set the read and write consistency levels always to be one. This means when there is a write operation, that 'write' must be written to the 'commit log' and 'memtable' of at least one replica node whereas when it is a read, it returns a response from the closest replica, as determined by the snitch. With the fixed settings of replication factor and consistency level we evaluated performance matrices of different models we proposed. As a matter of fact, we have set the operation count as 40,000 for each evaluation and all experiments ran with the Zipfian access distribution.

The base measurements we obtained from YCSB results were the system overall throughput (given as the average number of operations executed by the database per second), read latency and the write latency (given as the average response time taken for each operation). In general, for an any kind of database system, having higher throughput with a lower latency is always better.

\subsection{Measuring the Performance Overhead}
Initially, it was specifically focused on evaluating the performance overhead formed due to the cryptographic operations compared to the base model with no encryption. For this experiment we employed the EncM1 model in contrast to the NoEnc model. We evaluated each model against YCSB workload-a by increasing the number of client threads in a non-liner form 1, 2, 4, 8, 16, … up until 512. Each of configuration was performed ten times and the average was taken.

\begin{figure}[ht]
	\centering
	\includegraphics[width=0.99\columnwidth]{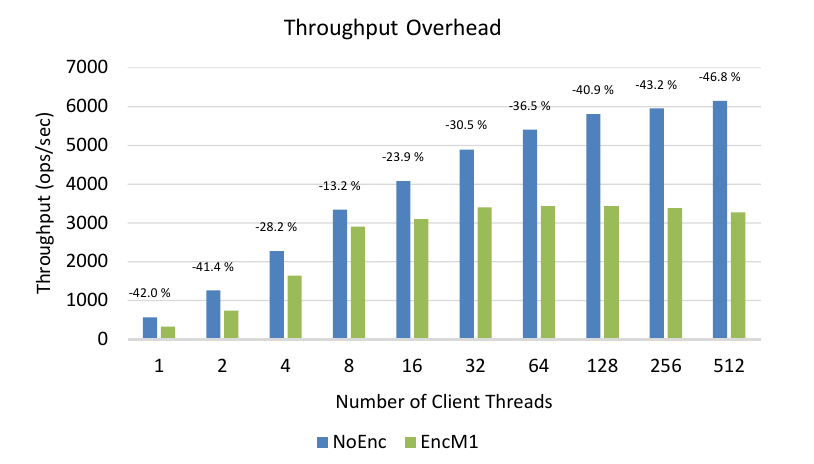}
	\caption{{Throughput Comparison of NoEnc and EncM1 Models.}}
	\label{fig:throughput_comparison}
\end{figure}

\begin{figure}[ht]
	\centering
	\includegraphics[width=0.99\columnwidth]{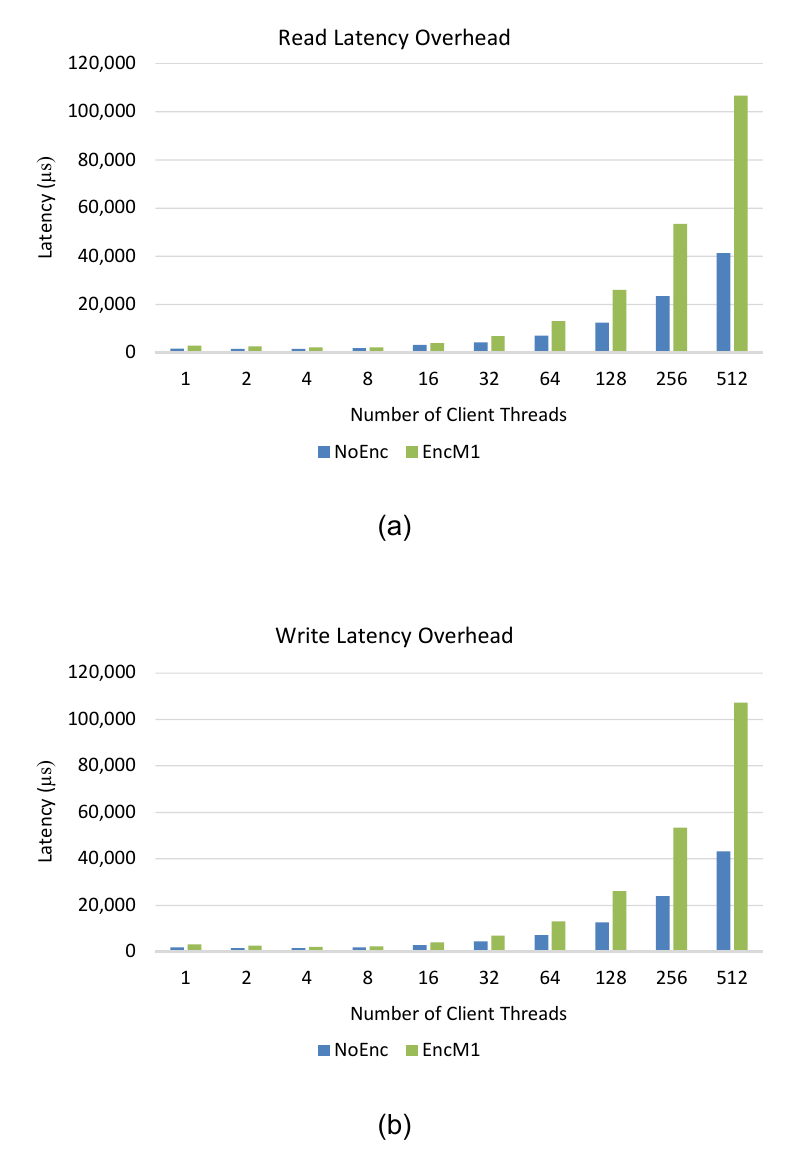}
	\caption{{Read and Write Latency Comparison for NoEnc and EncM1 Models.}}
	\label{fig:latency_comparison}
\end{figure}

Based on the results of our first experiment as shown in Fig. \ref{fig:throughput_comparison} \& Fig. \ref{fig:latency_comparison}, when there is only one client running, overall throughput of model EncM1 is decreased by 42\% while average read and write latencies are increased by 79\% and 73\% respectively. As the number of clients increases, the throughput overhead percentage is gradually decreases along with the decreasing read and write latency overhead percentages. This is simply because that the Proxy server can process multiple clients simultaneously. However, after some point, system throughput tends to be increased again and finally saturated around 3400 ops/sec. The reason behind the saturation point is the resource limitations in the Proxy server. As the number of clients are increasing more and more, Proxy server exceeds the maximum capacity that it can handle, result in more latency overhead.

\subsection{Performance Comparison of Different Models}
\begin{figure}[ht]
	\centering
	\includegraphics[width=0.99\columnwidth]{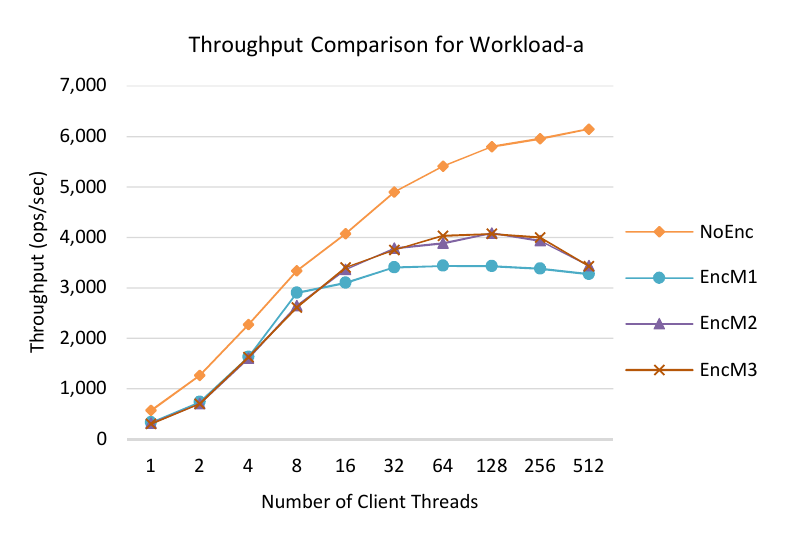}
	\caption{{Throughput Comparison of Different Models for Workload-a.}}
	\label{fig:throughput_diff_models}
\end{figure}

We then established an experiment to compare the throughput, read latency and write latency of different models that we proposed. Similar to the previous experiment with EncM1, we have chosen the YCSB workload-a which is defined as mix of 50\% read operations and 50\% write operations. Experiment was done by increasing the number of clients in a non-linear form 1, 2, 4, 8, … until 512. As the same way we conducted the first experiment, each model was evaluated by performing ten times and taking the average of them. Fig. \ref{fig:throughput_diff_models} shows the comparison of overall throughput of different models.

Since it does not apply any cryptographic operations, NoEnc model always has the best throughput among all. EncM1 is having a similar relationship with both EncM2 and EncM3 until number of clients exceeds four. However, thereafter EncM1 is having a slow increase in throughput and finally when the number of clients exceeds 32 it becomes more saturated. Comparatively, as it shows in Fig. \ref{fig:throughput_diff_models}, EncM2 and EncM3 performs better than EncM1 with the increase of number of clients. Thus, it is shown that adding more Proxy servers to the system to distribute the workload, system throughput can be increased.

Furthermore, it is also noted that both EncM2 and EncM3 are having almost the same identical pattern of throughput. The main reason behind this is, any Cassandra node can be the coordinator of the cluster and, coordinator can distribute the workload among other nodes. Therefore, either the Proxy servers are specifically connecting to a single node or connecting to all the nodes at the same time, does not have a much difference. In a different context, yet relevant to the same, after some point around 256 clients, all these models become saturated where they reach the maximum possible throughput on the allocated resources of the machine, for the given workload.

\begin{figure}[ht]
	\centering
	\includegraphics[width=0.99\columnwidth]{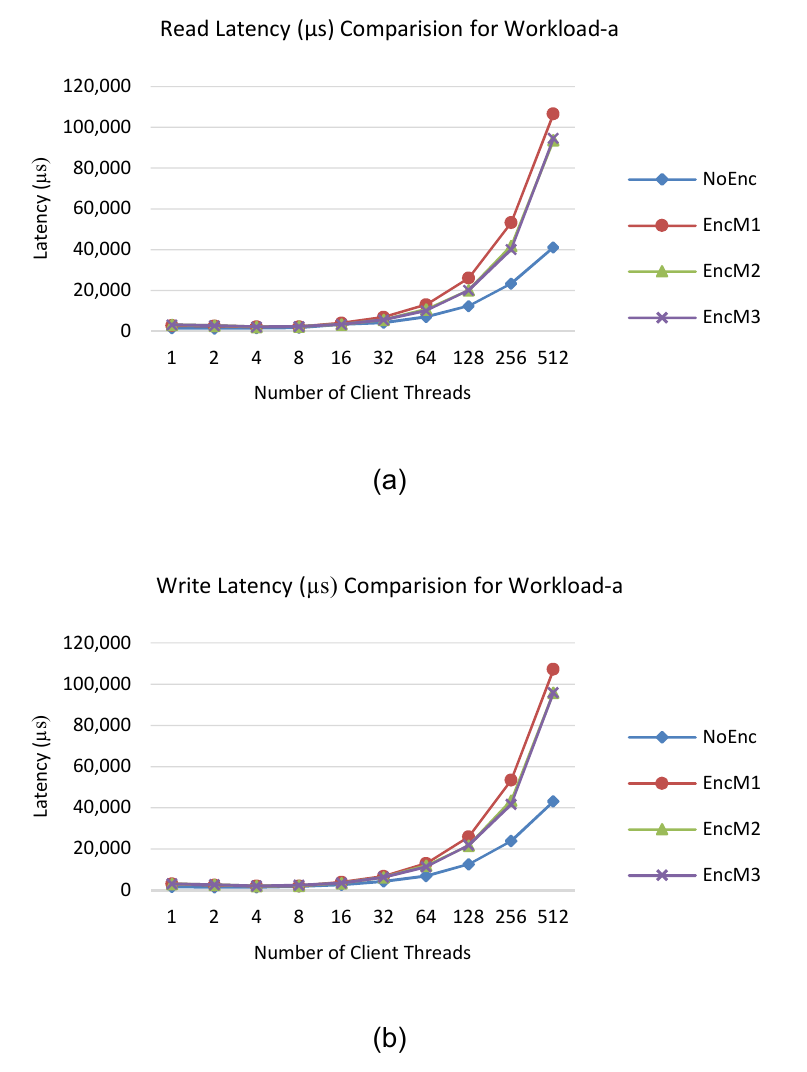}
	\caption{{Read and Write Latency Comparison of Different Models for Workload-a.}}
	\label{fig:latency_diff_models}
\end{figure}

Fig. \ref{fig:latency_diff_models} shows the comparison of different models against the latency. One of the important facts it reveals is, there is no much difference between read latency and write latency for a given model. At very early stage releases of Cassandra, it was reported that read operation latency is much higher compared to the write latency \cite{klein2015performance}, \cite{waage2014benchmarking}. Even though still there is a latency difference, we do not observe any major difference between the read and write latencies in the same model.

In a summary, with the addition of more Proxy servers to the system, overall performance can be improved. Also, our design performs better in a way compared to some of the related work done in the past, given in the literature. However, there is always a trade-off between cost of implementation and the performance. When the number of clients connected to the system is high, then it is good to have multiple Proxies, but it comes with an additional cost. At the same time, if the number of clients connected are less, it is shown that EncM1 might be a better selection as it does not have additional cost of implementation and it gives the same performance as others. However, with the proposed SEC-NoSQL deployment model, configuration and horizontal-scaling process of Proxy servers and database nodes is simplified.

\subsection{Effect of Having Different Workloads}
\begin{figure}[ht]
	\centering
	\includegraphics[width=0.99\columnwidth]{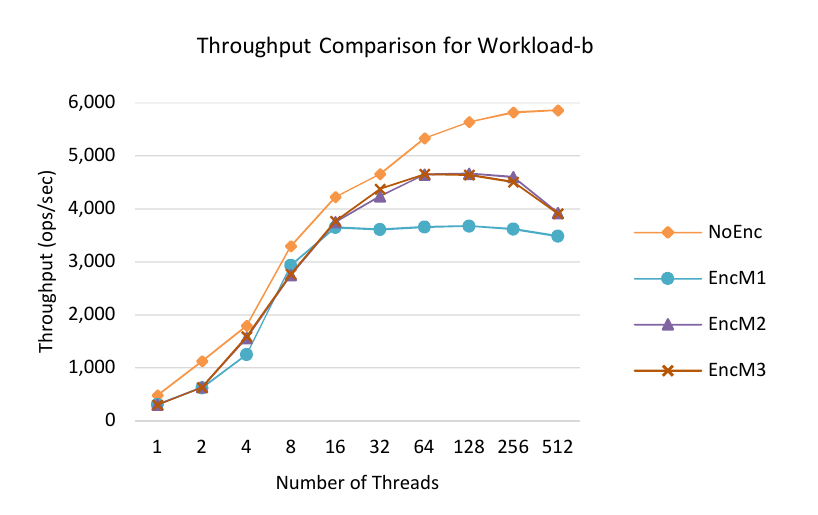}
	\caption{Throughput Comparison of Different Models for workload consists of 95\% read operations.}
	\label{fig:throughput_diff_models_b}
\end{figure}

\begin{figure}[ht]
	\centering
	\includegraphics[width=0.99\columnwidth]{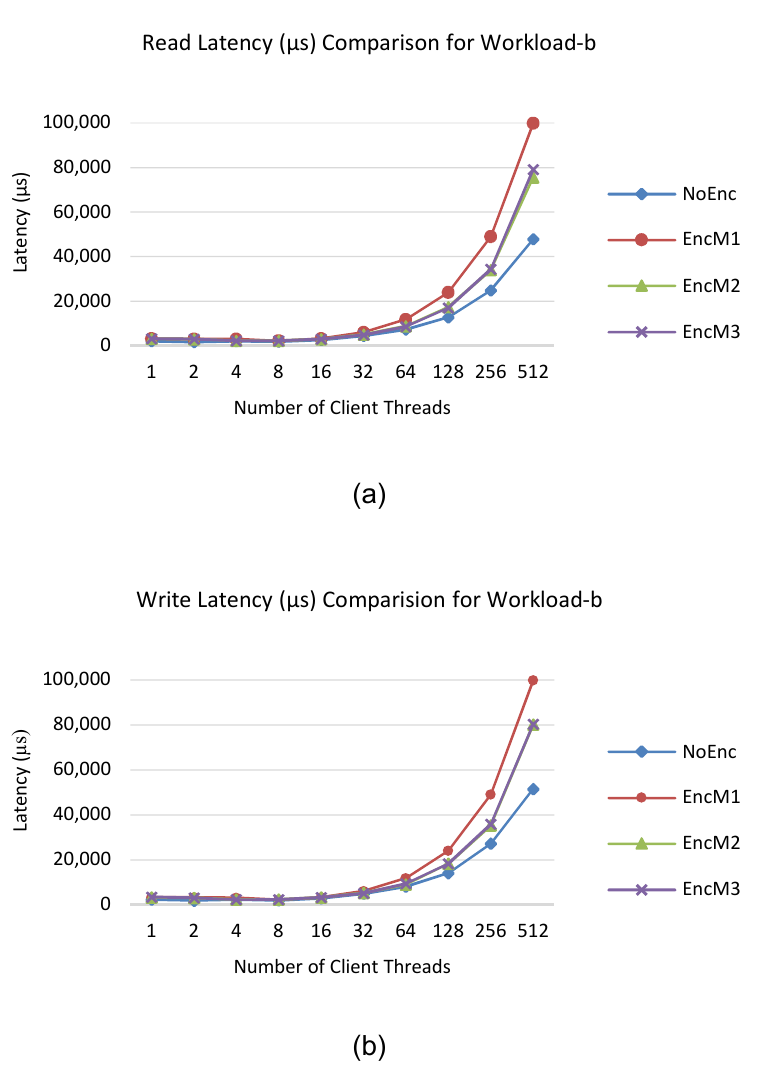}
	\caption{Read and Write Latency Comparison of Different Models for Workload-b.}
	\label{fig:latency_diff_models_b}
\end{figure}

Thereafter, we evaluated our models against a different workload (workload-b) by changing the proportion of read operations. We increased the read proportion to 95\% (earlier it was 50\%) by making the balance 5\% as write operations. As Cassandra read operations have higher latency than write operations, running a workload with high read portion, it is generally expected to notice a drop in overall throughput. Fig. \ref{fig:throughput_diff_models_b} shows the comparison of different models against throughput for the workload-b and Fig. \ref{fig:latency_diff_models_b} shows the latency comparison for the same. However, it is noticeable that even though there is a slight drop in overall throughput of NoEnc model, all other models have achieved better throughput compared to workload-a, after the number of clients exceeded 16. We attribute this to several factors. First, as we observed in the previous experiment as well, even though Cassandra read latency is much higher than write latency, the difference is minimal. Given that, when a request with a write query goes from Client to Proxy, number of bytes it takes through the communication channel is much higher compared to the request for a read. Thus, having higher portion of writes in a high volume of data make an additional overhead to the Proxy whereas higher portion of reads gives a performance improvement.

It is worth to note here that both in workload-a and workload-b, system tends to drop the performance after the number of clients exceeds 128. This could be mainly due to the resource limitations in the Proxy server(s). Finally, these series of experiments show that different workloads have different performance matrices, but regardless of the workload, deploying multiple proxy servers in a balanced order helps to improve the overall performance (bounded by the given resource allocations).

\subsection{Security-as-a-Service for NoSQL Databases}
When number of clients increases, the overall performance of SEC-NoSQL can be maintained at a guaranteed level by adding more Proxy servers to the system, certainly with additional costs. In this paradigm, SEC-NoSQL can be easily fitted in to a cloud-based architecture which can be operated in the model of Database-as-a-Service (DBaaS). The main advantage of having such a model is the flexibility and scalability. We moved one step further by coupling security with DBaaS model to make it a Security-as-a-Service model for NoSQL databases. The importance of having such solution is, users can still configure security based on their application specific requirements even without changing the existing client code/application or database engine. On the other hand, by having a cloud-based architecture, not only we can plug-in more database nodes, we can also plug-in more Proxy servers when it needs to provide guaranteed level of performance while providing security.

\begin{figure}[ht]
	\centering
	\includegraphics[width=0.99\columnwidth]{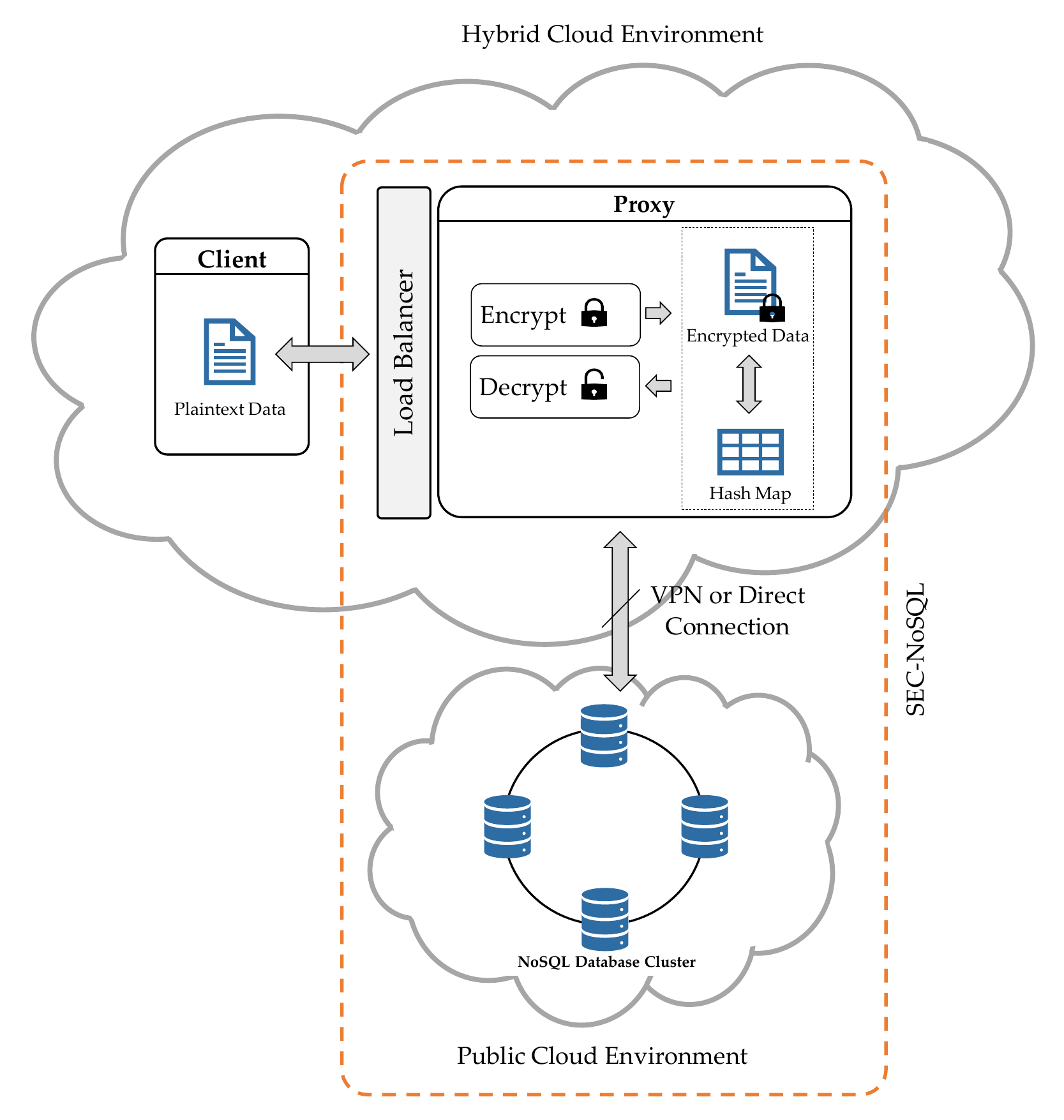}
	\caption{Deployment Architecture of SEC-NoSQL.}
	\label{fig:deployment_architecture}
\end{figure}

In SECaaS for NoSQL model, clients can enjoy all the inherent features of cloud-based database service, and it also ensures the confidentiality and integrity of the sensitive data. Since clients only communicate with the Proxy server(s) hosted on privately maintained cloud (or Hybrid cloud), they have the full control over their data and level of configuration. Fig. \ref{fig:deployment_architecture} shows high level deployment architecture of SEC-NoSQL.

To facilitate such a solution to the user, it is required to have a well-defined Service Level Agreement (SLA) between cloud service provider and the user. We define SEC-NoSQL SLA as follows. Suppose cloud service user is having number of clients $n \leq N$ accessing the service simultaneously. By operating $p\in P$ proxy servers in the system, service provider can facilitate;

\begin{enumerate}
	\item Guaranteed average throughput up to $T ops/sec$
	\item Average read latency not more than $l_r (s)$
	\item Average write latency not more than $l_w (s)$
\end{enumerate}

where $N$ and $P$ denotes the maximum number of clients and total number of Proxy servers operate by the service provider, respectively.

In a practical implementation scenario of our solution as a secure cloud service, given the values of $n$ and $p$, service provider should be able to properly estimate the guaranteed values of $T$, $l_r$ and $l_w$. Based on the performance evaluation results of this study, cloud service provider can generate some appropriate statistical models for throughput and latency along with the varying number of Proxy servers and clients. Based on these generated models, service provider can offer different set of service offerings to the user at a guaranteed level of throughput. Following example elaborate this idea in detail.

\begin{figure}[ht]
	\centering
	\includegraphics[width=0.99\columnwidth]{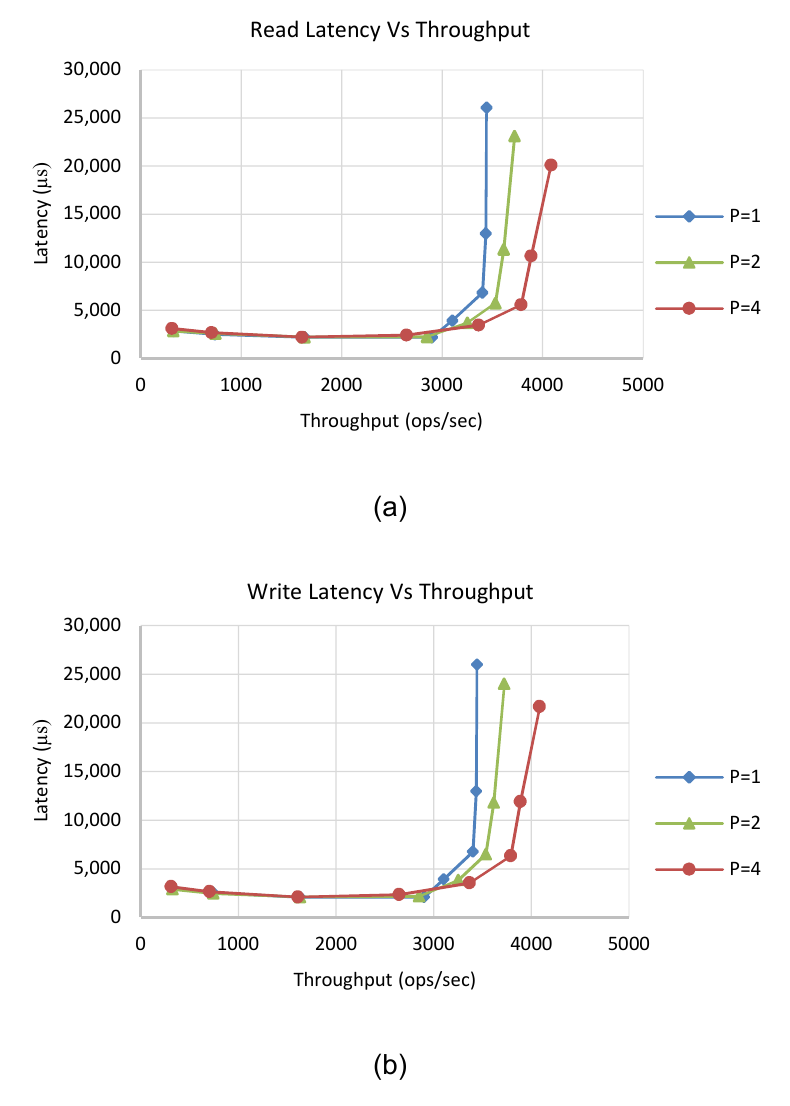}
	\caption{Latency Vs Throughput Evaluation Results.}
	\label{fig:latency_throughput}
\end{figure}

Let us assume that cloud service provider operates the system with $n$ client threads and $p$ Proxy servers where $n\leq 128$ and $p \in {1,2,4}$. Fig. \ref{fig:latency_throughput} shows the evaluation results for latency against throughput for different number of clients and Proxy servers. Then, based on these experimental results, a statistical model can be generated for each performance metric with respect to $n$ and $p$. We have modeled the extracted values as a polynomial surface in Matlab using data fitting tools however statistical models such as liner interpolation or exponential also can be considered.

Based on the generated liner model ‘Poly23’ each performance metric throughput, read latency and write latency can be formulated in to the following format of equation as in (1).

\begin{equation}
\begin{split}
f(p,n)=c_{00}+c_{10}p+c_{01}n+c_{20}p^2+c_{11}pn \\+c_{02}n^2+c_{21}p^2n+c_{12}pn^2+c_{03}n^3
\end{split}
\end{equation}

In the above equation, $c_{ij}$ are the coefficients derived from the performance curves obtained from the experiments. Using the generalized form of above equation, we can derive separate additional equations for throughput function as $T(p,n)$, read latency function as $l_r (p,n)$ and finally write latency function as $l_w (p,n)$.

Using the derived equations and the evaluation results, we have generated a system performance reference indicator table as an example shown in Table \ref{tab:sla_table}. For a given number of client connections, cloud service provider can offer three different options with separate service packages where each package has different level of guaranteed performance.

\begin{table}[hbtp]
	\centering
	\caption{SEC-NOSQL SLA}
	\includegraphics[width=0.99\linewidth]{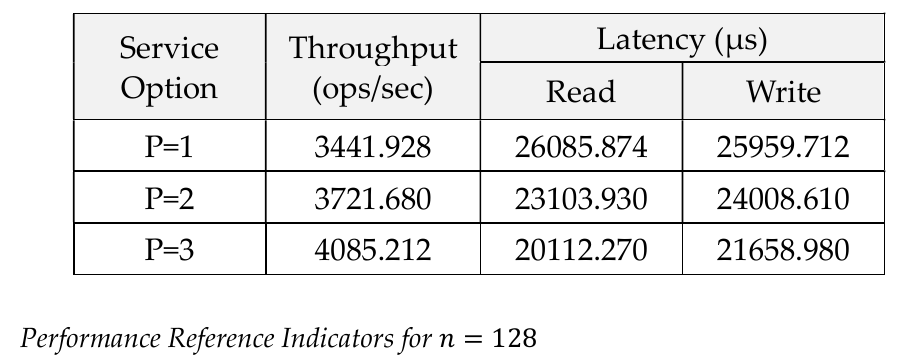}	
	\label{tab:sla_table}
\end{table}

For an example, if the user selects P=2 option, then service provider guarantees that user can experience the system throughput up to 3721.680 ops/sec while maximum latency of 23103.930\textmu s for read operations and maximum latency of 24008.610\textmu s for the write operations. Even though the values we have obtained in all these experiments are more implementation specific and need to have much more details like network infrastructure and cloud space allocations to establish a well-defined SLA, we believe that this piece of information can help to establish a cloud-based secure service with SEC-NoSQL. Moreover, as a future extension, we plan to extend this concept through a machine learning model (e.g. SVM) based on the experimental results to generate more accurate performance curves.

Apart from the technical discussions/implementations, it is also important to capture the different security requirements of the customers. For an example, there can be some users who have the strong technical potential to manage security for their demanding applications while some may not. Hence, it is better if the cloud service provider can facilitate different service packages for different user requirements regardless of users competency on security configurations. In addition, nowadays most could operators facilitate Pay-As-You-Go (PAYG) method which allows a user to scale, customize and provision computing resources. To that end, SEC-NoSQL can also be configured for different service packages along with PAYG method based on customer's requirement, their technical competence and investment towards cloud resources. So that, even if the customer is not technically sound to integrate security with hybrid-cloud deployments, they can still use SEC-NoSQL with customized security parameters by selecting the appropriate service package.

\section{Conclusion}
This paper takes the initiative towards implementing practical approach for high-performance, security-aware NoSQL databases which supports query over encrypted data deployed as a cloud-based service which supports Security-as-a-Service model for NoSQL databases. SEC-NoSQL is a practical design and implementation that can be easily adopted to existing NoSQL database environments without making any architectural changes to the client-side or database engine. Usually encrypting the data on a database affects the overall performance. We adopted state-of-the-art horizontal-scaling capabilities in a cloud environment to provide guaranteed level of database performance without much compromise even for high volume of client requests. We have implemented the complete solution on top of a latest Cassandra database cluster and elevated the performance of the proposed solution with various realistic workloads. To the best of our knowledge, none of previous work has designed such a system for NoSQL databases to deploy under a Security-as-a-Service model. The proposed three different implementation models with different configurations of SEC-NoSQL, were evaluated using YCSB. The results show that the user has the full flexibility to choose the best appropriate model based on his workload and the performance requirements. In addition, the first step towards providing SEC-NoSQL as a cloud service along with a framework for establishing a SLA is also provided.


%



\ifCLASSOPTIONcompsoc
\else
\fi


\ifCLASSOPTIONcaptionsoff
  \newpage
\fi




\bibliography{technical}
\bibliographystyle{ieeetr}
%


%







\end{document}